\documentclass[aps,prb,twocolumn,showpacs]{revtex4}

\usepackage{graphicx}

\providecommand\nima{Nucl.\ Instrum.\ Methods Phys.\ Res.\ A }
\providecommand\jap{J.\ Appl.\ Phys.\ }
\providecommand\sst{Supercond.\ Sci.\ Technol.\ }

\begin{document}

\title{Imaging of single infrared, optical, and ultraviolet
photons using distributed tunnel junction readout 
on superconducting absorbers}

\author{Miha Furlan, Eugenie Kirk, and Alex Zehnder}
\affiliation{Laboratory for Astrophysics, Paul Scherrer Institute,
5232 Villigen PSI, Switzerland}

\date{\today}

\begin{abstract}
Single-photon imaging spectrometers
of high quantum efficiency in the infrared to ultraviolet
wavelength range, with good timing resolution 
and with a vanishing dark count rate
are on top of the wish list in earth-bound astronomy, material and
medical sciences, or quantum information technologies.
We review and present improved operation of a 
cryogenic detector system potentially offering 
all these qualities. It is based on a superconducting absorber
strip read out with superconducting tunnel junctions. 
The detector performance is discussed in terms of responsivity,
noise properties, energy and position resolution.
Dynamic processes involved in the signal creation and detection
are investigated for a basic understanding of the physics, and
for possible application-specific modifications of 
device characteristics.
    
\end{abstract}

\pacs{07.60.Rd, 42.79.Pw, 85.25.Oj}

\maketitle


\section{Introduction\label{intro.sec}}
Superconducting tunnel junction (STJ) detectors are among the
most advanced cryogenic sensors\cite{Booth1996} with intrinsic 
spectroscopic resolution and high detection efficiency over a 
broad energy range. Essential advantages of low-temperature 
detectors in general 
are the small energy gap in superconductors compared to
standard semiconductors (typically three orders of magnitude lower),
decoupling of the electron and phonon systems 
in normal metals below about $50\, \mathrm{mK}$,
and a vanishing lattice heat capacity.

The lifetime of nonequilibrium quasiparticles generated 
in superconductors due to local energy deposition 
in excess of a purely thermal distribution 
becomes very large at operating
temperatures well below the transition temperature. 
Those quasiparticles can be efficiently detected as a tunnel current
in an STJ if tunneling rates are high compared
to recombination and loss processes.
In the tunneling process, a quasiparticle can tunnel back and 
contribute many times to the total charge transfer.\cite{Gray1978}
This internal amplification process is a unique feature of
STJs which significantly improves the sensitivity of the detectors,
allowing us to observe single sub-eV energy quanta.

As a next generation detector, 
STJs are very promising candidates e.g.\ for 
astronomical observations.\cite{Perryman2001,Martin2004}
In addition to their direct spectroscopic response and high speed
($\lesssim 10^5\, \mathrm{counts/s}$), they can be operated 
with essentially no dark count rate, which is another enormous
advantage compared to CCD imaging systems, making them the favored
choice for detection of faint objects.
However, to satisfy the desire for detectors with
imaging capabilities as offered by CCD cameras,
a competitive STJ based multipixel camera inevitably faces the 
complexity of large channel-number readout, i.e.\ 
the problem of transferring 
the small charge-signals from the cryogenic environment to external 
room-temperature electronics. One possible solution to reduce 
the number of readout lines is by using distributed readout 
imaging devices (DROIDs), where the quasiparticle distribution
in a large superconducting absorber is detected with several
STJs at different positions. This type of detector
was investigated with x-rays%
\cite{Kraus1989,Jochum1993,Friedrich1997,Kirk2000,Li2000,Hartog2000,Li2003} 
as well as with optical photons.%
\cite{Wilson2000,Verhoeve2000,Jerjen2006,Hijmering2006a,Hijmering2006b}

In this paper we present measurements with quasi-onedimensional
DROIDs irradiated with energy quanta covering the ultraviolet (UV)
and the entire optical range, and including  an 
extension to near infrared (IR) energies below $1\, \mathrm{eV}$.
After a description of device properties, 
experimental conditions and measurement procedure 
in Sec.~\ref{experiment.sec}, the linearity of detector response,
the noise performance and the resolution in energy and position
are discussed in Sec.~\ref{response.sec}.
Time-dependent processes in the detector involving quasiparticle
diffusion, loss and trapping rates, are considered in more detail
in Sec.~\ref{diffusion.sec}. 
The paper concludes in Sec.~\ref{conclude.sec}.


\section{Experiment\label{experiment.sec}}
The devices were fabricated by sputter deposition on a sapphire
substrate. The Ta absorber was deposited first, with an area of 
$L_\mathrm{Ta} \times W_\mathrm{Ta} = 135 \times 31.5\, \mu\mathrm{m}^2$
and a thickness of $100\, \mathrm{nm}$.
The square shaped Ta-Al junctions with a side length 
$L_j = 25\, \mu\mathrm{m}$ and an Al layer thickness of 
$38\, \mathrm{nm}$ were fabricated at each end on top of the absorber.
Hence, the length of the bare absorber was 
$L_\mathrm{b,Ta} = 80\, \mu\mathrm{m}$.
The residual resistivity ratio of the absorber was
$RRR = \rho_{_\mathrm{RT}} / \rho_{_\mathrm{LT}} \approx 25$, where
$\rho_{_\mathrm{RT}}$ and $\rho_{_\mathrm{LT}}$ are the (normal) 
resistivities at room temperature and low temperature, respectively.

The cryogenic detectors were operated at $310\, \mathrm{mK}$ and the 
supercurrent was suppressed by application of a magnetic field
parallel to the tunnel barrier.
With two independent charge-sensitive preamplifiers operated
at room temperature the devices were
voltage biased at $V_b = 80\, \mu\mathrm{V}$ where a thermal current
below $50\, \mathrm{pA}$ was measured. At this bias voltage the
current $I$ was found to depend on temperature $T$ according to
the empirical expression 
$I = 0.345 \exp(T / 26.4\, \mathrm{mK})\, \mathrm{fA}$.
Although the minimum thermal current was observed at 
$V_b \approx 200\, \mu\mathrm{V}$, the responsivity of the 
devices showed a maximum at lower $V_b$, in the range of
negative differential resistance of the current-voltage
characteristics.
We introduced ohmic resistors of $5\, \mathrm{k}\Omega$ in the signal 
lines (in series) close to the devices, which acted together with the 
junction capacitances as efficient $RC$ filters against high-frequency
noise on the $1.2\, \mathrm{m}$ cables, 
as well as damping resistors against resonances
of the junction-cable system corresponding to an $LRC$ circuit
with intrinsically low damping constant. 
Compared to the superconducting device resistance on the order 
of $1\, \mathrm{M}\Omega$ the influence of the filters on the 
detector response was negligible.

A variety of pulsed light-emitting diodes (LEDs) was used 
as light sources,  with wavelengths ranging from
IR at $1550\, \mathrm{nm}$ ($0.801\, \mathrm{eV}$) to 
UV at $370\, \mathrm{nm}$ ($3.355\, \mathrm{eV}$).
The specified spectral widths of the LEDs were typically a few percent.
The photons were coupled to the detectors via optical fibres 
and through the sapphire substrate. 
Back-illumination through the substrate was found to be essential
for optimum detector performance, probably thanks to a clean, 
oxide-free interface, whereas the native TaO$_\mathrm{x}$ on 
top of the absorber appears to be less transparent.
The light intensity (pulse duration)
was adjusted to observe on the order of one photon per pulse.
Packets of (incoherent) multiple photons were synchronized to a 
maximum time spread of less than $200\, \mathrm{ns}$.

Detection efficiency of Ta absorbers is about $60\%$ for optical and
UV photons, but it drops drastically below red photon energies 
$\lesssim 2\, \mathrm{eV}$.\cite{Weaver1974} 
Radiation shorter than about $200\, \mathrm{nm}$ is expected to 
be cut off by the sapphire substrate.\cite{Gervais1991}
Unfortunately, we had no reference system for calibrating the
absolute photon absorption probability.

The electronic signals from the preamplifiers had a risetime of about
$1\, \mu\mathrm{s}$ (comparable to the tunneling time and
to the $\sim 1\, \mathrm{MHz}$ 
bandwidth of the readout electronics) and a decay on the order of
$30\, \mu\mathrm{s}$.
The preamplifier outputs were band-pass filtered with a 
$10\, \mu\mathrm{s}$ time constant and digitized for offline analysis.

Due to noticeable dependence of the device responsivity 
on $V_b$ and on electronics settings the system was calibrated against 
an electronic pulser injecting a defined number of charges.
The charge noise of the readout electronics with the junctions connected
(but not irradiated) was determined from pulser distribution
spread or from rms  noise measurements to be
$q_e \approx 2760\, \mathrm{e}$ for both channels combined,
i.e.\ about $1950\, \mathrm{e}$ for each individual channel 
(at $V_b = 80\, \mu\mathrm{V}$).

\begin{figure}
\includegraphics[width=.98\columnwidth]{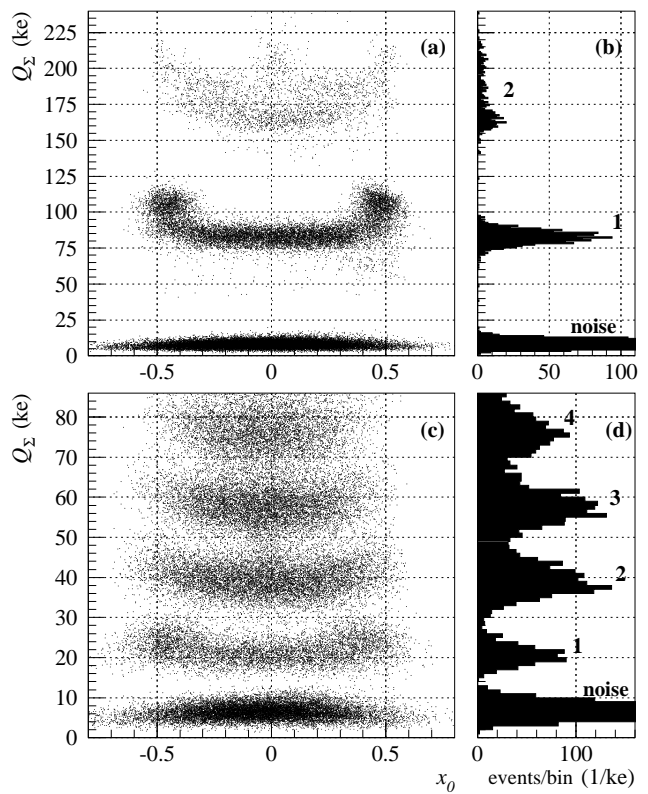}
\caption{\label{uv_ir.f}%
(a)~Total charge output $Q_\Sigma$ versus normalized 
position for the DROID 
irradiated with $3.355\, \mathrm{eV}$ (370 nm) UV photons.
(b)~Projection of the charge spectrum (histogram) of the
data in (a) for $| x_0 | < 0.05$.
The labels to the right of the peaks refer to the number of 
simultaneous photons, noise-triggered events are labelled with `noise'.
(c),(d)~Data as presented in (a),(b), but with the detector
irradiated with $0.801\, \mathrm{eV}$ (1550 nm) IR photons.}
\end{figure}

Figure~\ref{uv_ir.f} shows results from measurements with UV as well
as with IR photons. The charges $Q_1$ and $Q_2$ refer to the 
separate outputs from
the two channels. The left-hand plots display the total charge 
$Q_\Sigma = Q_1 + Q_2$ versus the normalized (onedimensional) position
of the photon interaction $x_0 = (Q_1 - Q_2) / Q_\Sigma$. 
Slight asymmetries of the data with respect to $x_0$ are due to
the difficulty of perfect adjustment of operating point 
for the two channels.
The shape of the event distribution is determined (in first order) 
by quasiparticle diffusion, loss processes, local energy gap,
and the efficiencies of trapping and tunneling 
(see Sec.~\ref{diffusion.sec}).

Spectral response and resolution of the experimental data
are determined for events with 
$| x_0 | < 0.05$, 
a projection of which is shown in the right-hand plots of 
Fig.~\ref{uv_ir.f}. The number of simultaneously 
absorbed energy quanta 
is given to the right of the corresponding peaks, 
whereas `noise' denotes events triggered on photonless
fluctuations.
A clear  spectral separation of single-photon events from 
multiple photons and from noise triggers is observed, 
down to the lowest measured IR energy quanta of $0.801\, \mathrm{eV}$


\section{Discussion\label{discuss.sec}}
As a first remark, we wish to comment on the events observed
in Fig.~\ref{uv_ir.f}(a) in the range
$(0.3 \lesssim x_0 \lesssim 0.6)$ and
$(55\, \mathrm{ke} \lesssim Q_\Sigma \lesssim 75\, \mathrm{ke})$.
They originate from single-photon energy depositions in the ground lead
connecting to the absorber. This lead is attached asymmetrically
to the detector, 
very close to junction 1. Therefore, a fraction of the 
quasiparticles generated in the ground electrode leak into 
that STJ and contribute a measurable signal.
This effect exemplifies that one can in principle design
any absorber shape appropriate for a specific imaging purpose.

The expected number $N_0$ of excited quasiparticles generated
upon deposition of energy $E$ is $N_0 = E / 1.7 \Delta$ where
$\Delta$ is the energy gap of the superconducting 
absorber and $1.7 \Delta$ is the effective energy to 
create one quasiparticle.\cite{Kurakado1982,Rando1992,Zehnder1995} 
Note that the energy conversion factor $1.7$ is appropriate
for Ta, but in general it is material dependent.\cite{Zehnder1995}
The intrinsic energy fluctuations $\varepsilon_i$ of STJs can 
be described in first order by\cite{Goldie1994}
\begin{equation}\label{enoise.eq}
\varepsilon_i = 
\sqrt{1.7 \Delta (F + 1 + \langle n \rangle^{-1} )}\, ,
\end{equation} 
where $F \approx 0.2$ is the Fano factor\cite{Kurakado1982,Rando1992}, 
and $\langle n \rangle$ is the average number each 
quasiparticle contributes to the charge signal due to 
backtunneling.\cite{Gray1978}
The energy-resolving power $R$ is conventionally described by
$R = E / 2.355 \varepsilon_i$ where the factor $2.355$ scales
between standard deviation and full width at half maximum of a
normal distribution.

\subsection{Linear response and spectral resolution\label{response.sec}}
At temperatures well below the superconducting transition
temperature,
where quasiparticle recombination processes are very slow, and 
for sufficiently low energy densities of the nonequilibrium 
quasiparticle distribution 
(i.e.\ for optical photons, where self-recombination
is negligible), a linear response of the detector is 
predicted for single STJ detectors\cite{Ivlev1991} as well as
for DROIDs.\cite{Esposito1994}
In order to test the linearity of our detectors we performed
measurements with photon energies ranging from IR to UV.
Similar to Figs.~\ref{uv_ir.f}(b) and (d) we consider only photon 
events with interactions in a narrow window of the 
absorber's central region, i.e.\ satisfying $| x_0 | < 0.05$.
In addition to single-photon events we also extract the signals
due to simultaneous two-photon events where data of minimum 
interaction distance is selected. This condition corresponds to the 
low edge of the two-photon events contour in Fig.~\ref{uv_ir.f}(a)
and to the prominent peak at $\sim 160\, \mathrm{ke}$ in
Fig.~\ref{uv_ir.f}(b), whereas 
events of two photons interacting at largest distance 
(i.e.\ close to the junctions) are expected to appear at the 
upper contour edge. 

\begin{figure}
\includegraphics[width=.98\columnwidth]{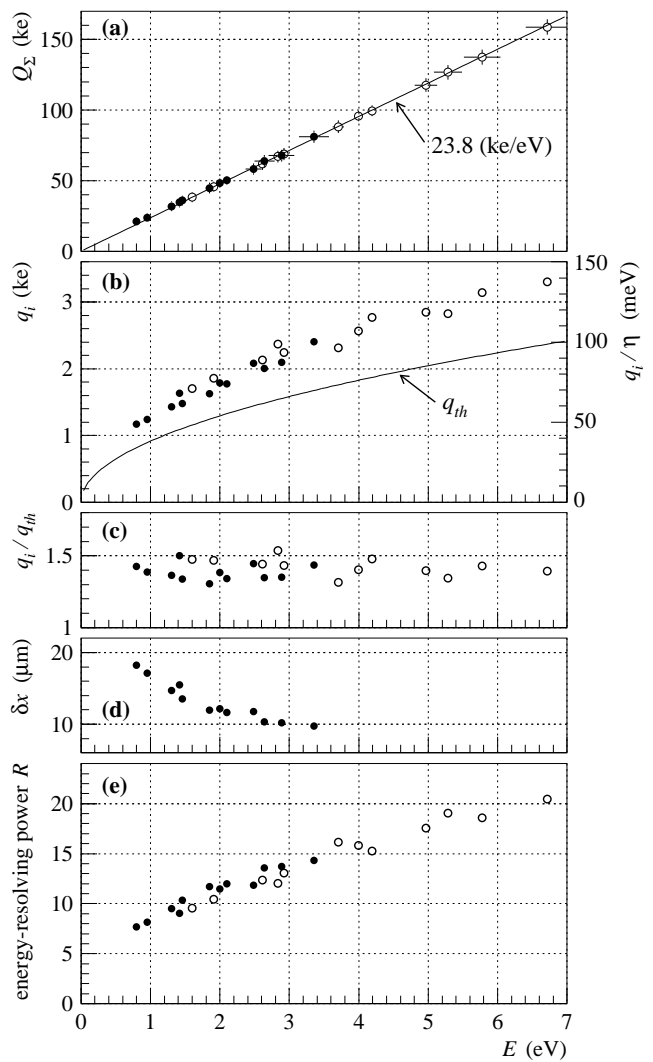}
\caption{\label{Q_vs_E.f}%
(a)~Detected total charge as a function of photon energy.
Full dots and open circles correspond to single and
two-photon events, respectively. The horizontal error bars are 
the specified spectral widths of the LEDs, the vertical errors
are taken from the experimental detector signal distributions.
The straight line is a least-square fit to the data points with
a slope as indicated.
(b)~Intrinsic experimental detector noise (electronic noise
and photon energy spread subtracted) versus energy. 
The drawn line is a theoretically predicted thermodynamic
limit. The right ordinate relates the data to units of eV.
(c)~Ratio of measured to theoretical intrinsic detector noise.
(d)~Estimated onedimensional position resolution.
(e)~Energy-resolving power $R = Q_\Sigma / 2.355 q_i$.}
\end{figure}

The results of detected total charge $Q_\Sigma$
as a function of photon energy are shown in Fig.~\ref{Q_vs_E.f}(a).
Solid dots and open circles correspond to single and two photons,
respectively. The horizontal error bars refer to the spectral
widths $\varepsilon_\lambda$ of the LEDs, 
the vertical errors reflect the measured
distribution spreads $q_m$ of signal amplitudes.
A linear least-square fit to the data points yields 
\[\eta \doteq Q_\Sigma / E = 23843\, \mathrm{e}/\mathrm{eV}.\] 
This is by a factor $28.4$ times more charge output than the 
theoretically generated $N_{0} = 840$ quasiparticles
per eV in a Ta absorber, where
$\Delta_\mathrm{Ta}=0.7\, \mathrm{meV}$. 
This amplification is attributed to 
the device-internal gain due to backtunneling.
In the case of lossless diffusion the responsivity would
even amount to $27.9\, \mathrm{ke}/\mathrm{eV}$ 
(see Sec.~\ref{diffusion.sec}),
corresponding to an average backtunneling factor 
$\langle n \rangle = 33.2$.

We found no significant deviations from linearity within experimental
errors over the entire measured energy range including the data from
two-photon events. This is most remarkable because it proves not
only the proportional regime considered in 
Refs.~\onlinecite{Ivlev1991,Esposito1994} but also an 
energy-independent backtunneling process at these 
quasiparticle densities.

By subtracting the readout electronics noise $q_e$ and the 
spectral widths $\varepsilon_\lambda$ of the LEDs from the 
measured pulse-height distribution spreads $q_m$ we deduce 
the device-intrinsic charge fluctuations $q_i$ as
$q_i^2 = q_m^2 - q_e^2 - (\varepsilon_\lambda \eta)^2$.
Figure~\ref{Q_vs_E.f}(b) shows $q_i$ as a function of 
energy  together with the theoretical fluctuations
$q_\mathrm{th} = \varepsilon_i \eta$ as given by 
Eq.~(\ref{enoise.eq}).
The ratio $q_i / q_\mathrm{th} \approx 1.4$ plotted in
Fig.~\ref{Q_vs_E.f}(c) suggests that the performance of
our detectors is close to the thermodynamic limit imposed by
the simple model~(\ref{enoise.eq}). The constant ratio implies
that further additive, energy-independent 
terms in Eq.~(\ref{enoise.eq}) may
better account for the device-intrinsic 
fluctuations.\cite{Segall2000,Martin2006}

If we neglect quasiparticle losses during diffusion,
the charge noise can be translated into an
imaging position resolution  $\delta x \approx 
(L_\mathrm{Ta} / x_\mathrm{max}) (q_i / Q_\Sigma)$
as displayed in Fig.~\ref{Q_vs_E.f}(d), where
$x_\mathrm{max} = 0.41$ in our devices accounts for 
the reduced trapping efficiency (see Sec.~\ref{diffusion.sec}).
The energy-dependent number of effective pixels 
$ \sim L_\mathrm{Ta} / 2.355 \delta x$ is in our case
in the range of $3 \ldots 6$ (neglecting the real geometry
where about one third of the absorber is modified by 
trapping regions). A possible improvement in 
position resolution, however, can be obtained
for longer absorbers (for the price of
reduced energy resolution due to diffusion losses) 
by measuring the time delay of the charge signals in the 
two channels. Assuming a moderate timing accuracy of
$1\, \mu\mathrm{s}$ we estimate from our timing measurements
(roughly $6\, \mu\mathrm{s}$ differential delay 
over $L_\mathrm{b,Ta}$, see Sec.~\ref{diffusion.sec})
a spatial resolution of $\sim 6$ pixels,
which is already comparable to the charge-noise limited values.

Finally, the energy-resolving power $R$ of our DROIDs
is plotted in Fig.~\ref{Q_vs_E.f}(e), showing a comfortable
signal-to-noise ratio for single-photon imaging down to 
near IR wavelengths.

\subsection{Quasiparticle diffusion\label{diffusion.sec}}
After localized energy deposition in a superconductor, cascades
of several fast processes eventually end with a population of 
excess quasiparticles. Those initial conversion processes are 
negligibly fast compared to quasiparticle recombination and loss 
times.\cite{Rando1992,Kozorezov2000}
Therefore, we expect our measurable device dynamics to be
dominated by diffusion and subsequent trapping and tunneling
processes. Quasiparticle diffusion in the absorber can be described
by the onedimensional equation
\begin{equation}\label{diffuse.eq} 
\frac{\partial}{\partial t} n(x,t) -
D \frac{\partial^2}{\partial x^2} n(x,t) +
\frac{1}{\tau_\mathrm{loss}} n(x,t) = 0\, ,
\end{equation} 
where $n(x,t)$ is the quasiparticle density, $D$ is the
diffusion constant, and $\tau_\mathrm{loss}^{-1}$ is the
loss rate in the Ta absorber.
We adopt the phenomenological model in Ref.~\onlinecite{Jochum1993}
which derives the final integral of collected charges
$Q_{k=1,2}$ at the two STJs
\begin{equation}\label{intQ.eq} 
Q_k (x_0) = Q_0 \frac{\sinh \xi_k \alpha + 
                \beta \cosh \xi_k \alpha }
 {(1 + \beta^2) \sinh \alpha + 2 \beta \cosh \alpha}\, ,
\end{equation}  
where $\xi_k = \vert x_0 - (-1)^k \vert /2$ 
is a normalized photon position relative to junction $k$, 
the dimensionless parameter
$\alpha = L / \sqrt{D \tau_\mathrm{loss}}$ 
measures the quasiparticle diffusion length 
$\Lambda =  L / \alpha$ relative to the absorber length, and
$\beta = (\tau_\mathrm{trap} / \tau_\mathrm{loss})^{1/2}$ 
compares the trapping rate 
$\tau_\mathrm{trap}^{-1}$ to the loss rate.

\begin{figure}
\includegraphics[width=.98\columnwidth]{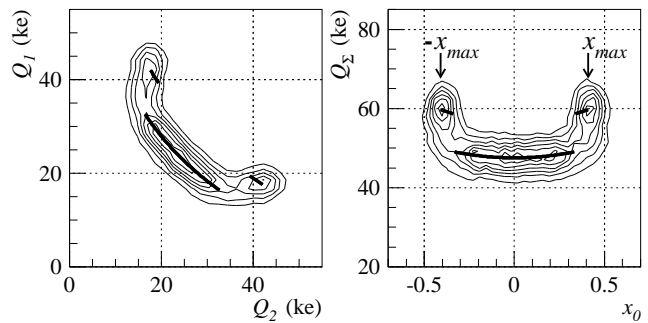}
\caption{\label{banana.f}%
Contour plots of a measurement with 
$\lambda = 592\, \mathrm{nm}$ photons (only single-photon
events are shown). The thick line represents the result of
a model fitted to the data points, where we 
assume onedimensional quasiparticle diffusion with losses
and a reduced gap in the trap region of the 
Ta absorber due to proximity of Al.
Photon absorption in the trap region is identified
with events at $x_0 = \pm x_\mathrm{max}$.}
\end{figure}

This model allows us to fit the experimental data as shown in
Fig.~\ref{banana.f}. In our specific devices, we have to distinguish
between the bare Ta absorber and the proximized
region due to trapping layers, where the effective gap energy relevant
for quasiparticle generation is reduced. Those regions are separated
at $| x_0 | \approx 0.3$. The position identified with energy 
depositions directly in the trap region is $x_\mathrm{max} = 0.41$
in our DROIDs.
Because $\alpha$ accounts for the curvature (losses) and $\beta$ for
the given $x_\mathrm{max}$ (trapping), we can fit
Eq.~(\ref{intQ.eq}) to the data in the limited range $| x_0 | < 0.3$,
yielding the fit parameters
$\alpha = 0.63$, $\beta = 0.57$ and $Q_{0,b} = 58.6\, \mathrm{ke}$
for the bare absorber.
Applying the same function to the data for $| x_0 | > 0.3$ while
keeping the formerly found $\alpha$ and $\beta$ fixed yields
$Q_{0,t} = 69.1\, \mathrm{ke}$ in the trap regions. 
This implies that the mean gap energy
in the trap region where quasiparticle generation takes place is
reduced by $(1-Q_{0,b}/Q_{0,t}) = 15.2 \%$. The empirical fit to
a Monte Carlo simulation\cite{Jerjen2006} has found a reduction
of about $12.5 \%$ for the same devices. 
The discrepancy between the results from the two approaches 
is acceptable considering the uncertainties in 
empirical parameter adjustment in the latter method.
One should note that the energy gap of 
$0.848 \Delta_\mathrm{Ta} = 594\, \mu\mathrm{eV}$
in the trap region is not equivalent to
the one at the tunnel barrier where about $450\, \mu\mathrm{eV}$
was found from current-voltage characteristics measurements.
This is well understood in terms of the superconducting proximity
effect.\cite{Zehnder1999}

For the parameter $\beta$ the simulations\cite{Jerjen2006} found
$\beta \approx (P_\mathrm{loss,Abs} / P_\mathrm{trap})^{1/2} = 0.483$.
This value is slightly lower than our model fit parameter because
the simulations\cite{Jerjen2006} consider a trapping and a tunneling 
probability separately (i.e.\ a nonvanishing detrapping probability), 
whereas our model averages over trapping and detrapping processes and
therefore results in a lower effective trapping rate.

The quasiparticle diffusion length in our absorber, 
i.e.\ the average length a quasiparticle travels before being lost 
by recombination or other loss channels, is
\[\Lambda =  \frac{L_\mathrm{Ta}}{\alpha} = 214\, \mu\mathrm{m},\]
which suggests that there is still room for a longitudinal
extension of the absorber.
In the expression above we use $L_\mathrm{Ta}$ 
(and not $L_\mathrm{b,Ta}$) as the 
effective absorber length based on the following argument:
A quasiparticle entering the trap region travels, 
if not tunneling before,  a distance $2 L_j$ until it has the 
first chance to detrap.  The average trapping length  
$\Lambda \beta^2$ is on the order of  $2 L_j$ (calculated with either
$L_\mathrm{Ta}$ or $L_\mathrm{b,Ta}$). The mean distance in 
the trap region travelled by quasiparticles which are being
trapped is therefore roughly $L_j$. Consequently, the absorber
length relevant for our analysis is 
$L_\mathrm{b,Ta} + 2 L_j = L_\mathrm{Ta}$.
Alternatively, we can consider quasiparticles starting in the 
trap region with arbitrary direction and position. Those which
are detected in the opposite STJ travel in average a distance 
$L_j$ before leaving the trap. Therefore, the events at 
$| x_\mathrm{max} |$ in Fig.~\ref{banana.f} are identified
with the physical positions at $\pm L_\mathrm{Ta}/2$, 
and not with the edge of the transition from bare absorber 
to trap region  at $\pm L_\mathrm{b,Ta}/2$.

\begin{figure}
\includegraphics[width=.98\columnwidth]{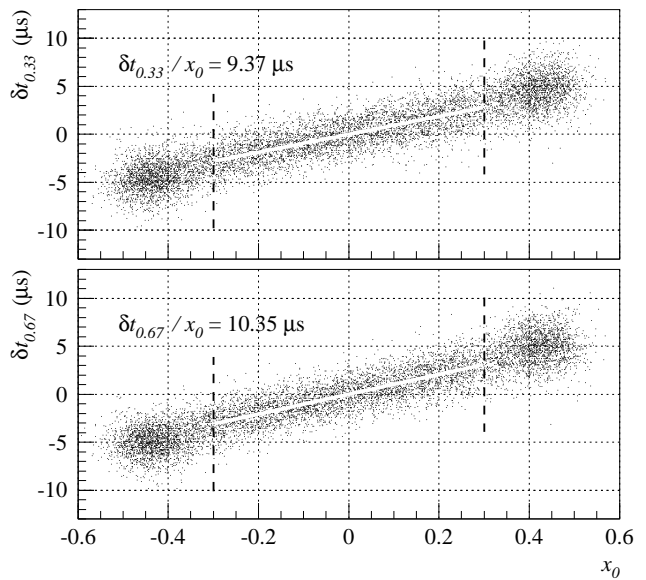}
\caption{\label{timing.f}%
Measured time differences between the two detector channels 
versus position, for trigger thresholds at 33\% (top) and 
at 67\% (bottom) of the signal peak amplitudes.
The white lines are linear fits to the data in the range
$| x_0 | < 0.3$. The resulting fit parameter is included
in the plots.}
\end{figure}

Measurements of  the trigger time differences $\delta t_m$ 
between the two
pulse-shaped detector signals for thresholds  set arbitrarily
at $33 \%$ and $67 \%$ of the peak amplitudes are shown in
Fig.~\ref{timing.f}. An S-shaped timing versus position
curve, which we observed
for fast signals, was stretched to a straight line after
pulse-shaping with a time constant one order of magnitude larger
than the original signal rise-time. The results of a linear 
fit to the data points in the range $|x_0| < 0.3$ are 
included in the figure.
The events corresponding to photon hits in the trap regions are
excluded from the fit because they show a slight extra time delay.

This observation is interpreted as a consequence of delayed
detrapping, namely the quasiparticles detected in the opposite
STJ have a non-zero probability to propagate within the trap 
for some time before leaving for the other side.
In addition, they may initially go through a 
few backtunneling cycles, which further delays the opposite
charge signal by the extra tunneling times.

We have numerically simulated the quasiparticle propagation according
to Eq.~(\ref{diffuse.eq}) to determine the detector-signal dynamics 
including the response of readout electronics and pulse shaping. 
The normalized timing differences $\delta t_n$ between the two 
channels calculated for trigger thresholds at $33 \%$ and $67 \%$
were $0.336$ and $0.351$, respectively, assuming $D=L=1$.
Scaling the measured data with these results allows us to estimate
the diffusion constant
\[D \approx \frac{\delta t_n}{2 \delta t_m\, x_\mathrm{max}} 
L_\mathrm{Ta}^2 \approx 7.8\, \frac{\mathrm{cm}^2}{\mathrm{s}}\, .\]
For comparison, a theoretically evaluated diffusion constant is given
by\cite{Narayanamurti1978}
\[D_\mathrm{th} = 
\sqrt{\frac{k_B T}{2 \pi \Delta}} v_{F}^2 \tau_\mathrm{imp}\, ,\]   
where $v_{F} = 1.6 \times 10^8\, \mathrm{cm}^2 \mathrm{s}^{-1}$
is the Fermi velocity of electrons in the superconductor,
$\tau_\mathrm{imp} = m_e / \rho_{_\mathrm{LT}} n e^2$ is the 
impurity-scattering time, $m_e$ is the electron rest mass and 
$n = 5.52 \times 10^{22}\, \mathrm{cm}^{-3}$ the density of
conduction electrons in Ta.\cite{Nussbaumer2000}
With $\rho_{_\mathrm{LT}} = 0.5 \times 10^{-8}\, \Omega\mathrm{m}$ 
(or $\rho_{_\mathrm{RT}} = 13 \times 10^{-8}\, \Omega\mathrm{m}$
and $RRR = 25$, respectively)
and  at $T = 310\, \mathrm{mK}$ we  obtain  
$D_\mathrm{th} = 25.2\, \mathrm{cm}^2 \mathrm{s}^{-1}$, which
is more than three times the experimentally determined
value. Similar discrepancies between experiment and
theoretical predictions were systematically observed in 
other experiments%
\cite{Friedrich1997,Kirk2000,Hartog2000,Nussbaumer2000,Li2003,Segall2004}
and remain unresolved.

The quasiparticle loss time deduced from the measurements
is found to be 
$\tau_\mathrm{loss} = \Lambda^2 D^{-1} 
\approx 58.9\, \mu\mathrm{s}$, which is significantly longer
than the tunneling time of about $1\, \mu\mathrm{s}$.
However, the trapping process with a time constant
$\tau_\mathrm{trap} = \beta^2 \tau_\mathrm{loss}
\approx 18.9\, \mu\mathrm{s}$ should preferably be improved 
in our devices towards faster trapping rates,
in order to enhance $x_\mathrm{max} \rightarrow 1$ for
better position resolution.

Alternatively to our phenomenological discussion on imaging
resolution in the previous section, the resolving capabilities
have been derived analytically\cite{Esposito1994} with a 
prediction for
\[ \delta x \approx \sqrt{D \tau_\mathrm{loss}} \frac{q_i}{Q_\sigma}
 = \Lambda \frac{q_i}{Q_\sigma}\, .\]
Inserting our experimental results (always for single-photon 
events only) yields position resolutions in the range
$\delta x \approx 26.7 \ldots 14.8\, \mu\mathrm{m}$, corresponding
to $5 \ldots 9$ pixels for our detector. This is in fair 
agreement with our former approximations.
However, we wish to emphasize that all these rough estimates 
ignore the geometrical properties of the real absorber,
which needs to be taken into account for a quantitatively
precise analysis of position resolution.\cite{Ejrnaes2005}


\section{Conclusions\label{conclude.sec}}
Detection of single (and simultaneous multiple) photons with good
spectral and spacial resolution was performed with STJ based
DROIDs. Single-photon resolution down to near IR energies
was proven, with a perfectly linear energy response over the entire
UV to IR wavelength range. 
Sensitivity to the investigated photon energies was possible
due to the backtunneling effect, delivering in our case a 
device-intrinsic gain of about 33, which is a feature unique to STJs.
The detectors were found to operate
close to the thermodynamic limit imposed by particle
fluctuation statistics.

The measured dynamic response of the DROIDs was compared to numerical
modelling based on quasiparticle diffusion including loss and trapping 
processes. The parameters extracted from experimental data 
are physically meaningful and coincide reasonably with 
former Monte Carlo simulations.\cite{Jerjen2006}
However, the diffusion constant was found to significantly disagree 
with theoretical predictions, similar to all preceding
experiments of the same kind. 

Position resolution of our relatively short absorber was 
estimated to about $3 \ldots 6$ equivalent pixels for photon
energies in the range  $0.8 \ldots 3.4\, \mathrm{eV}$.
By taking the differential time delays of the two channels
into account, we predict an improvement in position resolution 
by at least a  factor of two for a longer absorber 
($L_\mathrm{Ta} \rightarrow \Lambda$)  and better trapping 
efficiencies ($\tau_{trap} \ll \tau_{loss}$).
 
High-sensitivity spectrometers with single-photon counting capabilities
in the broad optical range are not only of high interest for
astronomical observations. 
Single photons at the telecommunication wavelength of 
$1550\, \mathrm{nm}$ are currently used, e.g., in intense studies
on quantum cryptography,\cite{Gisin2002} which links information 
theory to quantum entanglement physics.
Hence, STJ based DROIDs with the properties and potential as 
presented in this paper may as well be an interesting and
natural choice for IR single-photon  counting experiments.\\


\begin{acknowledgments}
We are grateful to Elmar Schmid for ceaseless improvement of the
readout electronics, to Iwan Jerjen and Philippe Lerch for valuable
discussions and experimental assistance, 
and to Fritz Burri for technical support.
\end{acknowledgments}






\begin{thebibliography}{99}

\bibitem{Booth1996}
N.~E.\ Booth and D.~J.\ Goldie, 
\sst \textbf{9}, 493 (1996).

\bibitem{Gray1978}
K.~E.\ Gray, \apl \textbf{32}, 392 (1978).

\bibitem{Perryman2001}
M.~A.~C.\ Perryman, M.\ Cropper, G.\ Ramsay, F.\ Favata, 
A.\ Peacock, N.\ Rando, and A.\ Reynolds,
Mon.\ Not.\ R.\ Astron.\ Soc.\ \textbf{324}, 899 (2001).

\bibitem{Martin2004}
D.~D.~E.\ Martin, P.\ Verhoeve, A.\ Peacock, A.\ van Dordrecht, 
J.\ Verveer, and R.\ Hijmering,
\nima \textbf{520}, 512 (2004).

\bibitem{Kraus1989}
H.\ Kraus, F.\ von Feilitzsch, J.\ Jochum, R.~L.\ M\"ossbauer,
Th.\ Peterreins, and F.\ Pr\"obst,
Phys.\ Lett.\ B \textbf{231}, 195 (1989).

\bibitem{Jochum1993} 
J.\ Jochum, H.\ Kraus, M.\ Gutsche, B.\ Kemmather, 
F.\ v.~Feilitzsch, and R.~L.\ M\"ossbauer,
Ann.\ Physik \textbf{2}, 611 (1993).

\bibitem{Friedrich1997}
S.\ Friedrich, K.\ Segall, M.~C.\ Gaidis, C.~M.\ Wilson, 
D.~E.\ Prober, A.~E.\ Szymkowiak, and S.~H.\ Moseley,
\apl \textbf{71}, 3901 (1997).

\bibitem{Kirk2000}
E.~C.\ Kirk, Ph.\ Lerch, J.\ Olsen, A.\ Zehnder, and H.~R.\ Ott,
\nima \textbf{444}, 201 (2000).

\bibitem{Li2000} 
L.\ Li, L.\ Frunzio, K.\ Segall, C.~M.\ Wilson, D.~E.\ Prober,
A.~E.\ Szymkowiak, and S.~H.\ Moseley,
\nima \textbf{444}, 228 (2000).

\bibitem{Hartog2000}
R.\ den Hartog, D.\ Martin, A.\ Kozorezov, P.\ Verhoeve, 
N.\ Rando, A.\ Peacock, G.\ Brammertz, M.\ Krumrey, 
D.~J.\ Goldie, and R.\ Venn,
Proc. SPIE \textbf{4012}, 237 (2000).

\bibitem{Li2003} 
L.\ Li, L.\ Frunzio, C.~M.\ Wilson, and D.~E.\ Prober,
\jap \textbf{93}, 1137 (2003).

\bibitem{Wilson2000} 
C.~M.\ Wilson, K.\ Segall, L.\ Frunzio, L.\ Li, D.~E.\ Prober,
D.\ Schiminovich, B.\ Mazin, C.\ Martin, and R.\ Vasquez,
\nima \textbf{444}, 449 (2000).

\bibitem{Verhoeve2000}
P.\ Verhoeve, R.\ den Hartog, D.\ Martin, N.\ Rando, 
A.\ Peacock, and D.~J.\ Goldie,
Proc. SPIE \textbf{4008}, 683 (2000).

\bibitem{Jerjen2006} 
I.\ Jerjen, E.\ Kirk, E.\ Schmid, and A.\ Zehnder, 
\nima \textbf{559}, 497 (2006).

\bibitem{Hijmering2006a} 
R.~A.\ Hijmering, P.\ Verhoeve, D.~D.~E.\ Martin, A.\ Peacock,
and A.~G.\ Kozorezov,
\nima \textbf{559}, 689 (2006).

\bibitem{Hijmering2006b} 
R.~A.\ Hijmering, P.\ Verhoeve, D.~D.~E.\ Martin, A.\ Peacock,
A.~G.\ Kozorezov, and R.\ Venn,
\nima \textbf{559}, 692 (2006).

\bibitem{Weaver1974}
J.~H.\ Weaver, D.~W.\ Lynch, and C.~G.\ Olson,
\prb \textbf{10}, 501 (1974).

\bibitem{Gervais1991}
F.\ Gervais, in \textit{Handbook of optical constants of solids II},
edited by E.~D.\ Palik (Academic, San Diego, CA, 1991), p.\ 761. 


\bibitem{Kurakado1982} 
M.\ Kurakado, 
Nucl.\ Instrum.\ Meth.\ \textbf{196}, 275 (1982).

\bibitem{Rando1992} 
N.\ Rando, A.\ Peacock, A.\ van Dordrecht, C.\ Foden, 
R.\ Engelhardt, B.~G.\ Taylor, P.\ Gare, 
J.\ Lumley, and C.\ Pereira,
\nima \textbf{313}, 173 (1992).

\bibitem{Zehnder1995}
A.\ Zehnder, \prb \textbf{52}, 12858 (1995).

\bibitem{Goldie1994}
D.~J.\ Goldie, P.~L.\ Brink, C.\ Patel, N.~E.\ Booth,
and G.~L.\ Salmon,
\apl \textbf{64}, 3169 (1994).

\bibitem{Ivlev1991}
B.\ Ivlev, G.\ Pepe, and U.\ Scotti di Uccio,
\nima \textbf{300}, 127 (1991).

\bibitem{Esposito1994}
E.\ Esposito, B.\ Ivlev, G.\ Pepe, and U.\ Scotti di Uccio,
\jap \textbf{76}, 1291 (1994).

\bibitem{Segall2000}
K.\ Segall, C.\ Wilson, L.\ Frunzio, L.\ Li, S.\ Friedrich,
M.~C.\ Gaidis, D.~E.\ Prober, A.~E.\ Szymkowiak, 
and S.~H.\ Moseley,
\apl \textbf{76}, 3998 (2000).

\bibitem{Martin2006}
D.~D.~E.\ Martin, P.\ Verhoeve, A.\ Peacock, A.~G.\ Kozorezov, 
J.~K.\ Wigmore, H.\ Rogalla, and R.\ Venn,
\apl \textbf{88}, 123510 (2006).

\bibitem{Kozorezov2000}
A.~G.\ Kozorezov, A.~F.\ Volkov, J.~K.\ Wigmore, A.\ Peacock,
A.\ Poelaert, and R.\ den Hartog,
\prb \textbf{61}, 11807 (2000).

\bibitem{Zehnder1999}
A.\ Zehnder, Ph.\ Lerch, S.~P.\ Zhao, Th.\ Nussbaumer, 
E.~C.\ Kirk, and H.~R.\ Ott,
\prb \textbf{59}, 8875 (1999).

\bibitem{Narayanamurti1978}
V.\ Narayanamurti, R.~C.\ Dynes, P.\ Hu, 
H.\ Smith, and W.~F.\ Brinkman,
\prb \textbf{18}, 6041 (1978).

\bibitem{Nussbaumer2000}
Th.\ Nussbaumer, Ph.\ Lerch, E.\ Kirk, A.\ Zehnder,
R.\ F\"uchslin, P.~F.\ Meier, and H.~R.\ Ott,
\prb \textbf{61}, 9719 (2000).

\bibitem{Segall2004}
K.\ Segall, C.\ Wilson, L.\ Li, L.\ Frunzio, S.\ Friedrich,
M.~C.\ Gaidis, and D.~E.\ Prober,
\prb \textbf{70}, 214520 (2004).

\bibitem{Ejrnaes2005}
M.\ Ejrnaes, C.\ Nappi, and R.\ Cristiano, 
\sst \textbf{18}, 953 (2005).

\bibitem{Gisin2002}
N.\ Gisin, G.\ Ribordy, W.\ Tittel, and H.\ Zbinden,
\rmp \textbf{74}, 145 (2002).







\end{thebibliography}
\end{document}